\documentclass[3p,times,procedia]{elsarticle}
\usepackage{nupha_ecrc}
\usepackage{color}

\volume{00}

\firstpage{1}

\journalname{Nuclear Physics A}

\runauth{}

\jid{nupha}

\jnltitlelogo{Nuclear Physics A}

\usepackage{amssymb}

\usepackage[figuresright]{rotating}

\begin{document}

\begin{frontmatter}



\dochead{}

\title{Initial state and pre-equilibrium effects in small systems}


\author{S\"{o}ren Schlichting}

\address{Physics Department, Brookhaven National Laboratory, Upton, NY 11973, USA}

\begin{abstract}
We discuss the importance of initial state and early time effects with regard to the theoretical understanding of long range azimuthal correlations observed in high-multiplicity $p+p$ and $p+A$ collisions.
\end{abstract}


\end{frontmatter}


\section{Introduction}
\label{seq:intro}
A surprising discovery in high-multiplicity proton-proton ($p+p$) and proton-nucleus $(p+A)$ collisions at the LHC was the observation of pronounced azimuthal correlations between produced particles which extend over several units in rapidity $\Delta\eta$ \cite{Khachatryan:2010gv,CMS:2012qk,Chatrchyan:2013nka,Aad:2012gla,Aad:2013fja,Aad:2014lta,Abelev:2012ola,Abelev:2013wsa}. While originally the study of $p+A$ collisions was designed as a reference measurement for nucleus-nucleus $(A+A)$ collisions to study so called 'cold nuclear matter' modifications of the initial state, the correlations observed in high-multiplicity events share many features reminiscent of $A+A$ collisions.

While in nucleus-nucleus collisions this behavior is attributed to the hydrodynamic expansion of a Quark-Gluon plasma, the theoretical interpretation of long-range correlations in small systems remains controversial  \cite{Dusling:2015gta}. Different theoretical explanations have been proposed based either on a final state response to the initial state geometry as in nucleus-nucleus collisions \cite{Bozek:2013uha,Bzdak:2014dia,Werner:2013ipa} or directly via correlations of particles produced in the initial state \cite{Dumitru:2010iy,Dusling:2012iga,Dusling:2012cg,Dusling:2012wy,Dusling:2013qoz,Dusling:2015rja}. However, it is presently still under debate whether the experimentally observed correlations in small systems originate pre-dominantly from initial state or final effects.

In this talk we address the question of multi-particle correlations in small systems from an initial state perspective. We provide a brief overview of the theoretical aspects of momentum space correlations in multi-particle production in Sec.~\ref{seq:theory}. Subsequently, in Secs.~\ref{seq::comparison} and \ref{seq:conclusion}  we critically assess the question to what extent experimentally observed correlations in $p+p$ and $p+A$ can be explained by initial state effects and outline further theoretical developments, which we believe are necessary to obtain a unified description of azimuthal correlations in small systems across a wide range of multiplicities.

\newpage
\begin{figure}[t!]
   \begin{center}   
       \includegraphics[width=\textwidth]{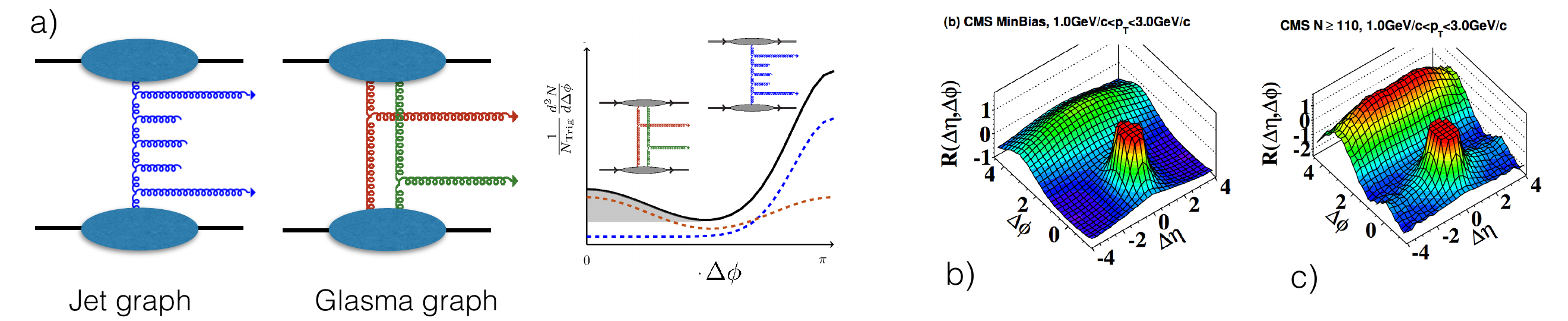} 

     \caption{ \label{fig:One} (Color online) a) Diagrammatic contributions to long range azimuthal correlations in high-energy scattering (from \cite{Dusling:2012cg}).\newline b \& c) Experimental result for di-hadron correlations in b) min. bias and c) high-multiplicity $p+p$ collisions at $7~\rm{TeV}$ (from \cite{Khachatryan:2010gv}).}
      \end{center}
\end{figure}

\section{Multi-particle production and Initial state correlations}
\label{seq:theory}
While multi-particle production in Quantum Chromo Dynamics (QCD) naturally leads to correlations between particles  produced in high-energy collisions, a complete theoretical understanding of these effects is extremely challenging. However, significant progress has been achieved in recent years based on the Color Glass Condensate (CGC) effective field theory of high-energy QCD \cite{Gelis:2008rw,Gelis:2008ad,Gelis:2008sz}, which provides the basis for phenomenological applications to hadronic collisions at RHIC and LHC energies \cite{Dumitru:2010iy,Dusling:2012iga,Dusling:2012cg,Dusling:2012wy,Dusling:2013qoz,Dusling:2015rja}. 

\subsection{Origins of long-range correlations}
By far the most famous example of initial state correlations is the back-to-back correlation between mini-jets or actual jets, which diagrammatically corresponds to the process shown in the left panel of Fig.~\ref{fig:One}a. In this case the production of multiple particles from a single hard scattering gives rise to an azimuthal correlation peaked around $\Delta \phi \sim \pi$ which is long range in the relative rapidity $\Delta \eta$ of the produced particles \cite{Dusling:2012cg}. Since 'jet-graphs' as in Fig.~\ref{fig:One}a dominate multi-particle production in min. bias or low-multiplicity events, such correlations are clearly seen in experimental studies of two-particle correlations in both $p+p$ and $p+A$ collisions. As an example Fig.~\ref{fig:One}b shows the CMS results \cite{Khachatryan:2010gv} for min. bias $p+p$ collisions where the away side ridge structure is clearly visible. 

When selecting events with high-multiplicity in the final state, it is important to realize that one is probing more exotic configurations of the hadronic wave functions which exhibit significantly larger parton densities. Consequently multi-particle production processes from multiple hard scatterings also shown in Fig.~\ref{fig:One}a become increasingly important with increasing multiplicity. Considering these 'Glasma graphs',  one finds that long range $(\Delta \eta)$ azimuthal correlations in the final state reflect the correlations of partons inside the hadronic wave-functions \cite{Dusling:2009ni,Kovchegov:2012nd,Altinoluk:2015uaa}. Within the color-glass condensate picture these correlations can be intuitively understood by considering the event-by-event fluctuations of chromo-electric fields inside the projectile and target \cite{Kovner:2010xk,Kovner:2012jm,Dumitru:2014dra,Dumitru:2014vka,Lappi:2015vta}. 

\subsection{Calculation of two-particle correlations}
\label{seq:PertCalc}
In practice, theoretical calculations of initial state two-particle correlations in $p+p$ and $p+A$  \cite{Dusling:2012iga,Dusling:2012cg,Dusling:2012wy,Dusling:2013qoz,Dusling:2015rja} are based on a direct computation of Glasma and Jet graphs in a $k_T$ factorized approximation valid for momenta above the saturation scale $Q_s$. While final state effects are ignored in these calculations, the effects of hadronization is included via fragmentation functions. Since interference effects vanish at leading order \cite{Dusling:2014oha}, the result is the direct sum of the Glasma and Jet graphs, yielding the following properties of the two-particle correlation function illustrated in the right panel of Fig.~\ref{fig:One}a. Besides the away side (mini-)jet contribution, Glasma graphs contribute a long range azimuthal correlation, peaked around $\Delta \phi \sim0$ and $\Delta \phi \sim \pi$ which is symmetric around $\Delta \phi=\pi/2$ and gives rise to the even harmonics $v_{2,4,\cdots}$ in the Fourier decomposition of the correlation function. Since the Glasma graph contribution increases with in the regime of high parton densities, its contribution is enhanced relative to the jet graph in high-multiplicity events, resulting in the emergence of an additional near-side $(\Delta \phi \sim 0)$ peak. Qualitatively, a similar trend can also be observed in the CMS data \cite{Khachatryan:2010gv} for high-multiplicity $p+p$ collisions ($N_{\rm{trk}}^{\rm{offline}}>110$) shown in Fig.~\ref{fig:One}c.

\begin{figure}[t!]
   \begin{center}   
       \includegraphics[width=\textwidth]{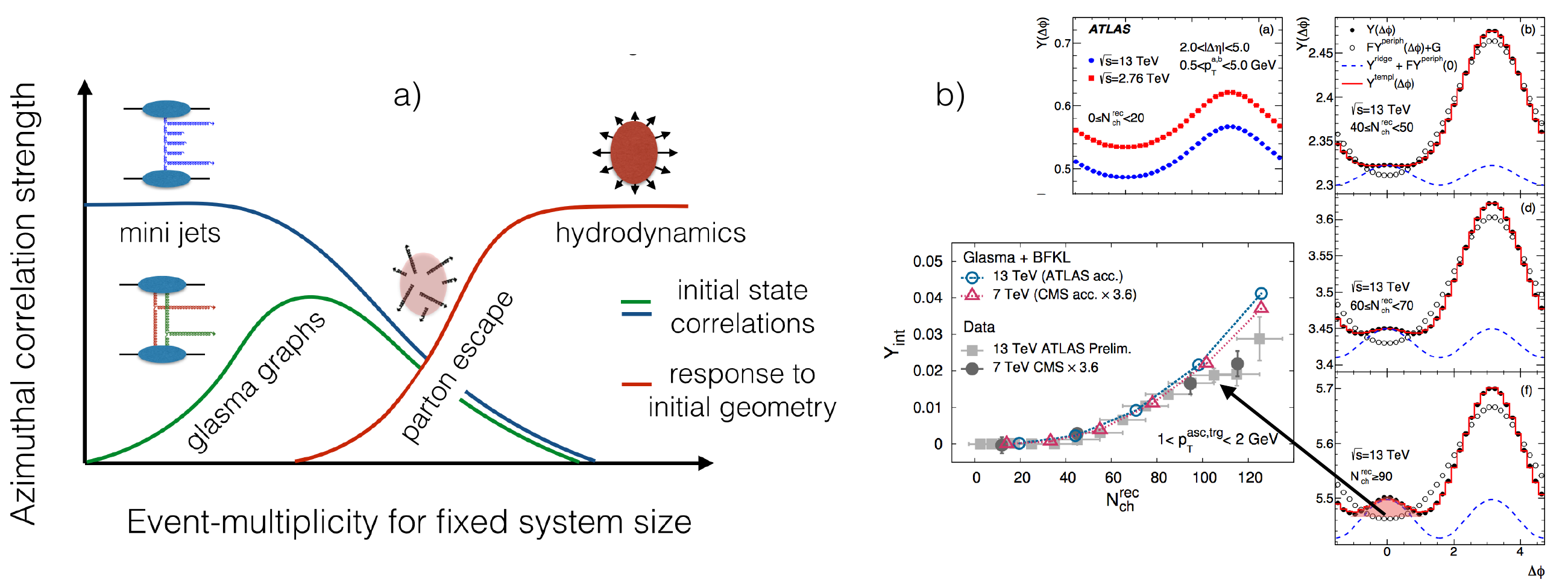} 

     \caption{ \label{fig:Two} (Color online) a) Phase diagram of long range azimuthal correlations in small systems. b) Experimental results for azimuthal correlations in p+p collisions \cite{Aad:2015gqa} and comparison of the near-side ridge yield to initial state calculations \cite{Dusling:2015rja}.}
      \end{center}
\end{figure}

\subsection{New theoretical developments}
Beyond the perturbative approach outlined above there have also been new theoretical developments to study initial state correlations in event-by-event simulations in classical Yang-Mills theory \cite{Schenke:2015aqa}. While these calculations naturally include the Glasma graphs, they extend the reach of perturbative calculations towards lower $p_T$ by consistently including multiple-scattering effects as well as coherent re-scattering in the final state. Since event-by-event simulations in classical Yang-Mills theory also allow for an improved treatment of the impact parameter dependence, they can be used in the future to systematically study initial state effects across different collision geometries e.g. in $p+A$,~$d+A$ and $3He+A$ collisions at RHIC \cite{Adare:2013piz,Adare:2015ctn,Adamczyk:2015xjc}.

Even though at present these calculations do not yet include jet graphs\footnote{Note that also the interference contribution between Glasma graphs and jet graphs no longer vanishes beyond leading order.} and hadronization effects -- and thus do not allow for a direct comparison with experimental data -- simulations in this regard have lead to new insight into the correlations at the parton level concerning in particular the dynamics of the correlations during the very early stages. While gluons are produced with a significant momentum space anisotropy at $\tau=0^{+}$, initially the two-particle correlation function is  symmetric around $\Delta \phi=\pi/2$ and features only even harmonics in accordance with the perturbative result \cite{Schenke:2015aqa}. However, including the effects of the classical Yang-Mills evolution up to $\tau=0.4 \rm{fm}/c$ leads to the build up of a sizable $v_{3}$ on the parton level, while the initial state $v_{2}$ remains intact  \cite{Schenke:2015aqa}.

\section{Comparison with experimental observations in $p+p$ and $p+A$ collisions}
\label{seq::comparison}
Before we turn to a more detailed comparison of initial state calculations to experimental data in $p+p$ and $p+A$ collision it is useful to briefly summarize the theoretical expectations.  While so far we have only addressed correlations due to initial state effects it is important to establish a clear understanding of the interplay between initial state and final state effects in small systems.

Our schematic expectation is illustrated in Fig.~\ref{fig:Two}a, where the azimuthal correlation strength due to initial state and final state effects is shown versus the event multiplicity for a fixed system size e.g. in $p+A$ collisions and transverse momentum range e.g. $1-3~\rm{GeV}$. Based on our discussion we expect that in low multiplicity or min. bias events azimuthal correlations between $1-3~\rm{GeV}$ particles are pre-dominantly due to back-to-back mini-jets. With increasing event-multiplicity the contribution from Glasma graphs becomes increasingly important resulting in the emergence of an additional near side ridge. When increasing the multiplicities even further, final state interactions  in this transverse momentum region can no longer be neglected at some point  and lead to a depletion of both 'jet-like' and 'glasma-like' correlations. At the same time the system starts to show a response to the initial state geometry, which in this low opacity region is presumably dominated by the path length dependence of the parton energy loss -- also referred to as parton escape mechanism \cite{He:2015hfa}. Ultimately, in the limit of very high multiplicities, mini-jets are fully quenched resulting in the formation of a thermalized medium and the complete loss of initial state correlations. In this high opacity regime, azimuthal correlations are dominated by the response to initial geometry described by a hydrodynamic expansion of a thermalized Quark-Gluon plasma.

While from a theoretical perspective we expect to observe a behavior as in Fig.~\ref{fig:Two}a, it is extremely difficult to predict the location of the transition from an initial state to a final state dominated evolution. In fact this question is closely related to our understanding of the thermalization process in nucleus-nucleus collisions, which despite significant progress in recent years \cite{Berges:2013eia,Kurkela:2015qoa} has posed an outstanding challenge to theorists. Hence, a more promising approach is to directly attempt an extraction of the boundaries between the different regimes through detailed comparisons of theory and experiment. As a first step in this direction, we will now analyze to what extent experimental observations can be consistently explained by initial state effects.

\subsection{p+p collisions}
We begin with a comparison in $p+p$ collisions at LHC energies, where the latest $13~\rm{TeV}$ ATLAS data \cite{Aad:2015gqa} is shown in Fig.~\ref{fig:Two}b. A striking feature that emerges at first sight is the fact that the away side jet-contribution is essentially unmodified across the entire range of multiplicities. Based on our discussion, we believe that this observation by itself provides strong evidence that a dominant fraction of initial state correlations survives even in the highest-multiplicity windows. One also observes that, in addition to the dominant jet-peak, a near-side ridge emerges and becomes more prominent with increasing multiplicity. Within the initial state framework, this is naturally understood as the contribution from Glasma graphs. A quantitative comparison of the near-side yield in $p+p$ collisions also shown in Fig.~\ref{fig:Two}b  yields a good agreement between initial state calculations \cite{Dusling:2015rja} and experimental data both at $7$ and $13$ TeV. The fact that the near-side yield is approximately energy independent is also explained naturally within this framework \cite{Dusling:2015rja}.
\subsection{p+A collisions}
While a systematic comparison of initial state calculations to experimental data in $p+Pb$ collisions yields a similar level of quantitative agreement of two-particle correlations for momenta $p_T > 1~\rm{GeV}$ \cite{Dusling:2012wy,Dusling:2013qoz}, there has been enormous progress on the experimental side to identify additional signatures of collective motion which could be indicative of the onset of significant final state effects. Even though some observables, such as e.g. mass ordering properties observed in correlations between identified hadrons \cite{Abelev:2013wsa} are not necessarily sensitive to the origin of the correlations, more promising directions including e.g. correlations between more than two particles have also been explored \cite{Khachatryan:2015waa,Abelev:2014mda}. One of the most striking observations in this regard is the sign change of the four-particle cumulant $c_{2}\{4\}$ observed around $N_{ch}(|\eta_{lab}|<1) \sim 60$ by ALICE \cite{Abelev:2014mda}. However, most of the recent measurements have focused on low $p_T$ observables where the perturbative calculations \cite{Dusling:2012iga,Dusling:2012cg,Dusling:2012wy,Dusling:2013qoz,Dusling:2015rja} outlined in Sec.~\ref{seq:PertCalc} do not necessarily apply, hence complicating the comparison between theory and experiment. 

Nevertheless, there have been first attempts to extend the theoretical framework to understand whether certain features of the low $p_T$ data such as the sign change of $c_{2}\{4\}$ can also be explained from initial state effects. While perturbatively $c_{2}\{4\}$ is found to be positive, it was argued in \cite{Dumitru:2014yza} that non-Gaussian correlations of color-electric fields inside the nucleus could indeed cause a sign change of the four particle cumulant. Similarly, preliminary results from event-by-event simulations in classical Yang-Mills theory a la \cite{Schenke:2015aqa} also suggest a negative sign of $c_{2}\{4\}$ when all particles have low momenta, while at high $p_T$ the four particle cumulant is always seen to be positive \cite{slides}. While the first results are interesting, further theoretical progress is needed to decide unambiguously whether the sign change in $c_2\{4\}$ can be plausibly explained in terms of initial state and early-time effects. It would also be interesting to extend the experimental measurements of higher cumulants towards higher momenta to achieve convergence on this issue.

\section{Conclusions \& Outlook}
\label{seq:conclusion}
Multi-particle production in QCD leads to long range azimuthal correlations of particles produced. While these initial state effects are small in nucleus-nucleus collisions, they can be sizable in small systems and should be taken into account in a consistent theoretical description. 

Calculations based purely on initial state correlations quantitatively describe the experimental data in $p+p$ collisions up to the highest multiplicity windows. Most importantly, the interpretation of the near-side ridge as an initial state effect is consistent with the fact that the away side jet peak is unmodified --  which by itself strongly constrains the magnitude of final state effects.

While a similar level of quantitative agreement is reached in comparisons of initial state calculations and experimental data for two-particles correlations in $p+Pb$ with momenta $p_T > 1~\rm{GeV}$  a simultaneous description of low and high $p_T$ data across a wide range of multiplicities remains challenging within any theoretical framework. Clearly, this points to the importance of developing a theoretical framework which consistently takes into account both initial state and final state effects.

With this challenge also comes the opportunity to explore novel aspects of the non-equilibrium QCD dynamics during the early stages of high-energy collision experiments. With additional information provided from $p+A$,~$d+A$ and $3He+A$ collisions at RHIC~\cite{Adare:2013piz,Adare:2015ctn,Adamczyk:2015xjc}, we are looking forward to exciting developments in the study of small collision systems.\\

\textit{Acknowledgements:} The author acknowledges valuable discussions with A.~Dumitru, K.~Dusling,  T.~Lappi, L.~D.~McLerran B.~Schenke, P.~Tribedy and R.~Venugopalan on this topic. SS is supported under DOE Contract No. DE-SC0012704 and gratefully acknowledges a Goldhaber Distinguished Fellowship from Brookhaven Science Associates.





\bibliographystyle{elsarticle-num}
\biboptions{sort&compress}



\end{document}